\documentclass{epl}

\title{
A heat pump at a molecular scale controlled by a mechanical force
}
\author{
        Naoko Nakagawa\inst{1} \and Teruhisa S. Komatsu\inst{2}
}
\institute{
  \inst{1} College of Science, 
        Ibaraki University, Mito 310-8512, Japan\\
  \inst{2} Department of Physics, Gakushuin University,
         Mejiro, Tokyo 171-8588, Japan 
}

\begin{document}
\maketitle

\begin{abstract}
We show that a mesoscopic system such as Feynman's ratchet may operate as a heat pump,
 and clarify a underlying physical picture.
We consider a system of a particle moving along an asymmetric periodic structure.
When put into a contact with two distinct heat baths of equal temperature,
the system transfers heat between two baths as the particle is dragged.
We examine Onsager relation for the heat flow and the particle flow,
and show that the reciprocity coefficient is a product of
 the characteristic heat and the diffusion constant of the particle.
The characteristic heat is the heat transfer between the baths
associated with a barrier-overcoming process.
Because of the correlation between the heat flow and the particle flow,
the system can work as a heat pump when the particle is dragged.
This pump is particularly effective at molecular scales where the energy barrier is of
the order of the thermal energy.
\end{abstract}

\section{Introduction}

In the last decade, physics and technology have moved more and more
toward the manipulation of mesoscopic objects.
Techniques of micro manipulation of molecules are expanding to open
new possibilities to future technology.
Although to control heat flow in mesoscopic scales may be among crucial issues,
 its study is still in a preliminary stage \cite{Peyrard_Casatti}.
Heat pumps working in nano-scale devices
 \cite{Humphrey_Linke,Feldmann_Kosloff}
 or mesoscopic machines  \cite{Jarzynski_Mazonka}
 are studied
 but the number of cases are still limited.
In this Letter, we study a class of mesoscopic systems working as a heat pump
and present a physical picture explaining how they work.
We start from a numerical study adopting one specific model, and
 demonstrate that a characteristic heat in a barrier-overcoming
process determines the ability of the heat pump and Onsager's reciprocity coefficients.
Then, we give a theoretical derivation of the reciprocity coefficients and
 show the generality of the results in a class of Feynman's ratchet \cite{Feynman}.
We present a simple and clear picture for the mechanism of heat pumps, which
 is expected to be useful in designing various mesoscopic heat pumps.

\section{Model}

\begin{figure}[t]
\begin{center}
\onefigure[scale=0.85]{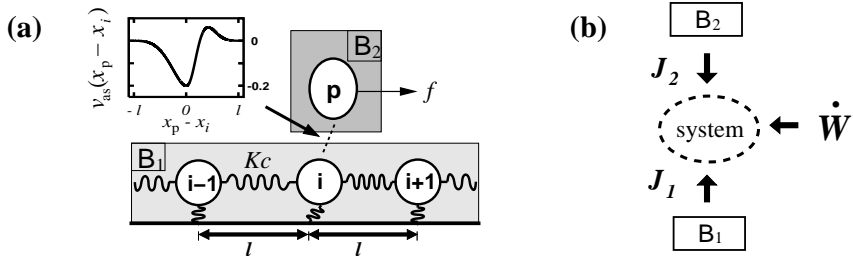}
\end{center}
\caption[]{
(a) Profile of the model for numerical simulation. 
We fix 
$\gamma\!=\!1$, $N_\mathrm{c}\!=\! 7$ with a periodic boundary condition,
and $K_\mathrm{c}=0.5$ except for Fig. \protect\ref{fig:Jq-Jp}.
The depth of $v_\mathrm{as}(x_\mathrm{p}-x_i)$ is $0.2$.
(b) The flow of energy in the system.
$J_1$ and $J_2$ are mean heat flows from heat baths $B_1$ and $B_2$
 to the system, and $\dot W$ is the mechanical work
 per unit time provided by the external force.
}
\label{fig:model}
\end{figure}

We study a class of systems including Feynman's ratchet in which a particle moves
 along a periodic structure.
The particle interacts with the periodic structure in an asymmetric manner.
The periodic structure and the particle
are attached to distinct heat baths
$B_1$ and $B_2$,
 whose temperatures are $T_1$ and $T_2$, respectively.
We start from a specific but typical model
 shown in Fig. \ref{fig:model} \cite{Nakagawa_Komatsu},
 which is suitable for numerical simulation.
In this model,
 a one-dimensional harmonic lattice corresponds to the periodic structure.
The time evolution of the system is described by
a set of Langevin equations,
\begin{equation}
\ddot x_i = -\gamma \dot x_i +\sqrt{2\gamma T_1} ~\xi_i(t)
-\frac{\partial (V_\mathrm{as}+ V_\mathrm{l})}{\partial x_i}, \quad
\ddot x_\mathrm{p} = -\gamma \dot x_\mathrm{p}
 +\sqrt{2\gamma T_2}~\xi_\mathrm{p}(t)
 -\frac{\partial V_\mathrm{as}}{\partial x_\mathrm{p}}+f,
\label{eqn:model}
\end{equation}
where  $x_\mathrm{p}$ and $x_i$ are the positions of the particle 
and the $i$-th lattice site ($i=1, \cdots, N_\mathrm{c}$).
An external force $f$ is applied to the particle.
$\gamma$ is the friction constant 
and $\xi_{\alpha}(t)$ is the random force
satisfying $\langle \xi_{\alpha}(t)\rangle=0$ and 
$\langle \xi_{\alpha}(t)\xi_{\beta}(t')\rangle=\delta_{\alpha,\beta}\delta(t-t')$.
The interaction between the particle and the lattice sites is described by 
 the asymmetric potential $V_\mathrm{as}=\sum_i v_\mathrm{as}(x_\mathrm{p}-x_i)$
 with $v_\mathrm{as}$ as in Fig. \ref{fig:model}.
$V_\mathrm{l}=\sum_i v_\mathrm{o}(x_i-x_{i0})+v_\mathrm{c}(x_{i+1}-x_{i})$ 
 is the harmonic potential for the lattice,
 where $v_\mathrm{o}(x)={K_\mathrm{c}}x^2/2$ is the on-site potential 
 and $v_\mathrm{c}(x)={K_\mathrm{c}}(x-l)^2/4$ is the interaction between neighbors.
$l$ is the lattice interval, $x_{i0}=il$ for $i$-th site,
 and $K_\mathrm{c}$ is a parameter for stiffness.
When $K_\mathrm{c}$ is sufficiently large,
the lattice does not fluctuate and
the potential for the particle reduces to
a one-dimensional sawtooth-shaped  potential.

On this periodic structure, the particle performs
 stepwise motions with a typical step size $l$.
At sufficiently low temperatures,
each stepwise motion occurs stochastically 
with a dwell time long enough for the particle to lose its memory.
In equilibrium, the probabilities of rightward and leftward steps
 are of course identical.
In the following, we concentrate on nonequilibrium steady states
 maintained by an external force
 and/or a temperature difference.

\section{Heat flow induced by particle motion and Onsager relation}
\begin{figure}[t]
\begin{center}
\onefigure[scale=0.9]{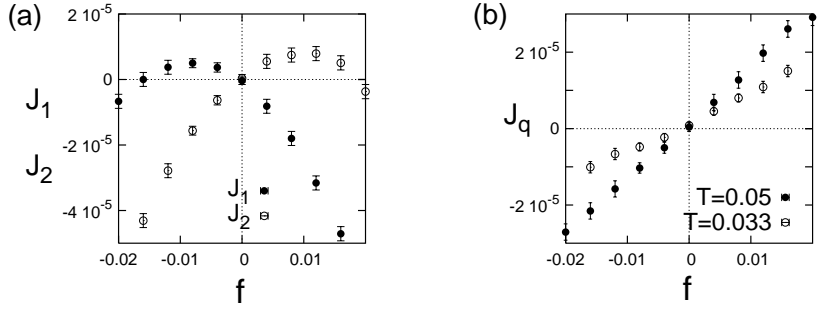}
\end{center}
\caption{
(a) Mean heat flows $J_1$ and $J_2$ as a function of
$f$ at $T_1=T_2=0.05$.
For small $f$, either $J_1$ or $J_2$ is positive,
 which indicates that the system absorbs energy from one of the baths.
(b) $J_\mathrm{q}$ vs $f$ for $T_1=T_2=0.05$ and $0.033$,
where $J_\mathrm{q} \propto f$. 
}
\label{fig:J1,J2}
\end{figure}

Temporal heat flows from the heat baths to the system
 are defined for each trajectory
 by using the stochastic energetics \cite{Sekimoto,Sekimoto_Takagi_Hondou} as
\begin{equation}
\hat{J}_\mathrm{1}(t) = \sum_i 
[ - \gamma \dot x_i +\sqrt{2\gamma T_\mathrm{1}}\,\xi_i(t) ]
 \circ \dot x_i, \qquad
\hat{J}_\mathrm{2}(t) =
 [ - \gamma \dot x_\mathrm{p}
 +\sqrt{2\gamma T_\mathrm{2}}\,\xi_\mathrm{p}(t) ]
 \circ \dot x_\mathrm{p},
\label{eqn:heatflow}
\end{equation}
where $\hat{J}_\mathrm{1}(t)$ is the heat flow from $B_1$ to the system
and $\hat{J}_\mathrm{2}(t)$ is that from $B_2$ to the system.
From Eq.(\ref{eqn:heatflow}),
 we define temporal heat flow between the two heat baths as
\begin{equation}
\hat J_\mathrm{q}(t)\equiv(\hat J_2(t)-\hat J_1(t))/2.
\end{equation}
In the steady states, mean heat flows 
$J_1 \equiv \langle \hat{J}_\mathrm{1}(t)\rangle$,
$J_2\equiv \langle \hat{J}_\mathrm{2}(t)\rangle$ 
and 
$J_\mathrm{q}\equiv \langle \hat J_\mathrm{q}(t)\rangle$
can be defined,
where $\langle\cdot\rangle$ denotes average over time and ensemble.

When a steady external force $f$ is applied to the particle,
energy corresponding to the mechanical work by $f$
is provided to the system.
Its mean rate is
\begin{equation}
\dot W \equiv \langle \dot x_\mathrm{p} f \rangle=J_\mathrm{p} f,
\label{eqn:work}
\end{equation}
where $J_\mathrm{p}$ is the mean particle flow defined as
$J_\mathrm{p}\equiv\langle \dot x_\mathrm{p} \rangle$.
In the steady states, 
the energy of the system remains constant on average.
This constraint leads to the balance of the energy flows 
$\dot W$, $J_1$ and $J_2$ as in Fig. \ref{eqn:model}(b),
 and implies $\dot W=-(J_1+J_2)$. 

Let $T_1=T_2$. When the particle is dragged,
 the particle and the sites are expected
 to warm up due to the friction.
Then one expects that heat transfer from the hot components
 to their surroundings takes place,
 and that $J_1<0$ and $J_2<0$. 
On the contrary to this expectation, 
$J_1$ or $J_2$ can be positive for small values of $f$ 
in Fig. \ref{fig:J1,J2}(a)
\footnote{
If the system is 
not in overdamped limit, 
the mechanical force can make some part of the system cool down.
When $J_2>0$, kinetic temperature of the particle is lower
than $T_2$ 
and, when $J_1>0$,
that for the lattice sites lower than $T_1$.
}.
The system then works as a heat pump which absorbs energy 
from one of the heat baths and dissipates it to the other.

Fig. \ref{fig:J1,J2}(b) shows that
 the heat flow $J_\mathrm{q}$ depends linearly on $f$ in a wide range,
 including the region with $J_1J_2>0$ (see Fig. \ref{fig:J1,J2}(a) ).
The particle flow $J_\mathrm{p}$ also shows a linear dependence on $f$
 in a comparable range.
Since the range in which the system works as a heat pump is included 
 in this linear range, we are led to study Onsager relation \cite{Onsager}
 for $J_\mathrm{p}$ and $J_\mathrm{q}$.
Define $T$ and $\Delta T$ by $T_1=T-\Delta T/2$ and $T_2=T+\Delta T/2$.
Thermodynamic forces for the flows $J_\mathrm{p}$ and $J_\mathrm{q}$
 are $f/T$ and $\Delta T/T^2$, respectively.
Onsager relations are formally written as,
\begin{equation}
J_\mathrm{p} =
{\displaystyle L_\mathrm{p p} \frac{f}{T}
 +L_\mathrm{p q}\frac{\Delta T}{T^2}},  \qquad
J_\mathrm{q} = 
{\displaystyle L_\mathrm{q p} \frac{f}{T}
 +L_\mathrm{q q}\frac{\Delta T}{T^2}}.
\label{eqn:reciprocal0}
\end{equation}
The coefficients are written as
\begin{equation}
L_\mathrm{pp}=D_\mathrm{p}, ~~L_\mathrm{qq}=T^2\lambda,
~~L_\mathrm{qp}=L_\mathrm{pq}=q^\mathrm{eq}D_\mathrm{p}/l,
\label{eqn:reciprocal}
\end{equation}
where $D_\mathrm{p}$ is the diffusion constant for the particle 
in equilibrium, $\lambda\equiv \partial J_\mathrm{q}/\partial \Delta T$ is
 the heat conductivity of the system and $q^\mathrm{eq}$ is
 the characteristic heat which will be discussed in the next section.
To derive the expression for $L_\mathrm{pp}$, we define the mobility as
 $\mu\equiv \partial J_\mathrm{p}/\partial f$.
Then Einstein's relation $\mu=D_\mathrm{p}/T$ implies
 $L_\mathrm{pp}=\mu T=D_\mathrm{p}$.
The expression for $L_\mathrm{qq}$ follows from the definition.
The expressions for the off-diagonal coefficients 
 $L_\mathrm{qp}$ and $L_\mathrm{pq}$
 will be derived in the following
 (see (\ref{eqn:chi_q-q}) and (\ref{eqn:Kubo3})).

\section{Characteristic heat associated with barrier-overcoming process}

\begin{figure}[t]
\begin{center}
\onefigure[scale=0.4]{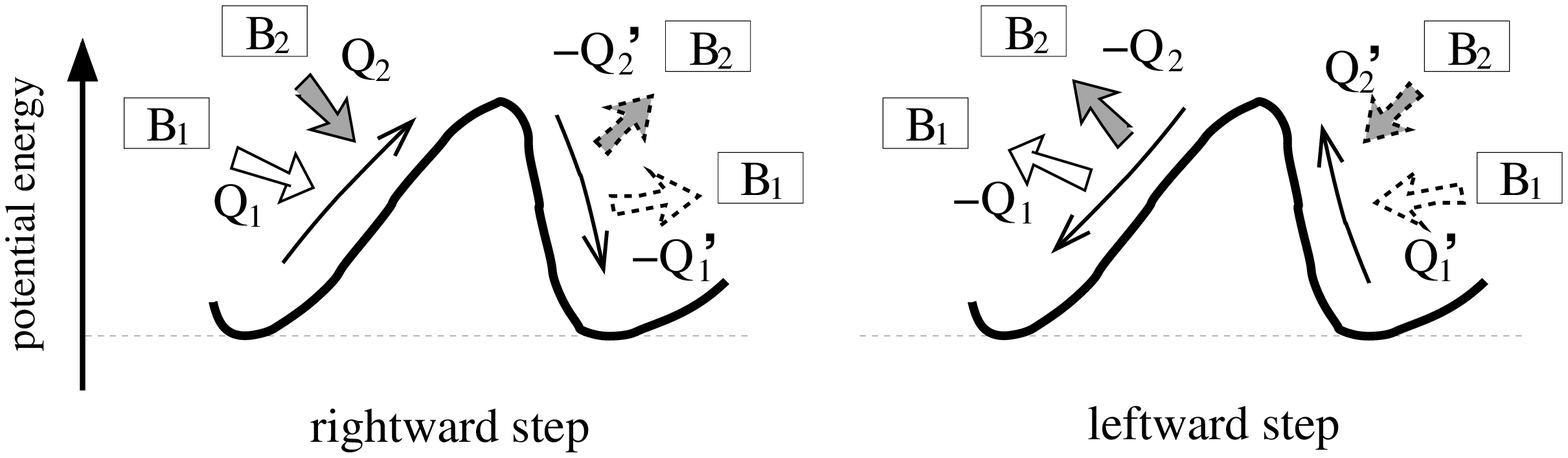}
\end{center}
\caption{
Schematics of energy flow when the system overcomes an energy barrier
in equilibrium.
The bright and dark arrows indicate the heat transfer from
 the heat bath $B_1$ and $B_2$, respectively.
In the rightward step, $Q_1$ is transfered from $B_1$ during the uphill
 and $-Q'_1$ during the downhill.
Because of the time-reversal symmetry in equilibrium states,
 $Q'_1$ is transfered from $B_1$ during the uphill
 and $-Q_1$ during the downhill in the leftward step.
Similar relations hold for $B_2$
 and the relation $Q_1+Q_2=Q'_1+Q'_2$ holds.
Because the system has an asymmetry,
 $Q_1\ne Q'_1$ and $Q_2\ne Q'_2$ in general.
}
\label{fig:qRqL-shematic}
\end{figure}

In this section, we investigate a characteristic heat associated with
 a barrier-overcoming process 
in the vicinity of the equilibrium states.
It enables us to determine
 the non-diagonal coefficients $L_\mathrm{qp}$ and $L_\mathrm{pq}$.
First, let us consider equilibrium cases.
When the system exhibits a stepwise motion,
 it should overcome an energy barrier.
To reach the top of the barrier
 the system absorbs some amount of heat from the two heat baths
 and dissipates the same amount of heat to descend from the top
 (Fig. \ref{fig:qRqL-shematic}).
In equilibrium, of course, no net heat transfer occurs
 between the baths on average.
But if we take the ensemble averages of the heat transfer for the 
 individual steps classifying the rightward and the leftward steps,
 nonvanishing heat transfer between the two baths can survive.
We find that a
 single rightward step carries heat $q_\mathrm{R}$ and
 a leftward step carries $q_\mathrm{L}$ on average. 
In equilibrium, the rightward and the leftward stepwise motions
 occur with the equal probability and
 the relation $q_\mathrm{R}=-q_\mathrm{L}\equiv q^\mathrm{eq}$
 holds, which consistently leads to vanishing of net heat transfer
 between the two heat baths.
In general, nonvanishing values of $q^\mathrm{eq}$ are obtained
 \cite{Komatsu_Nakagawa},
 because it comes from left-right asymmetry of the system.

\begin{figure}[t]
\begin{center}
\onefigure[scale=0.9]{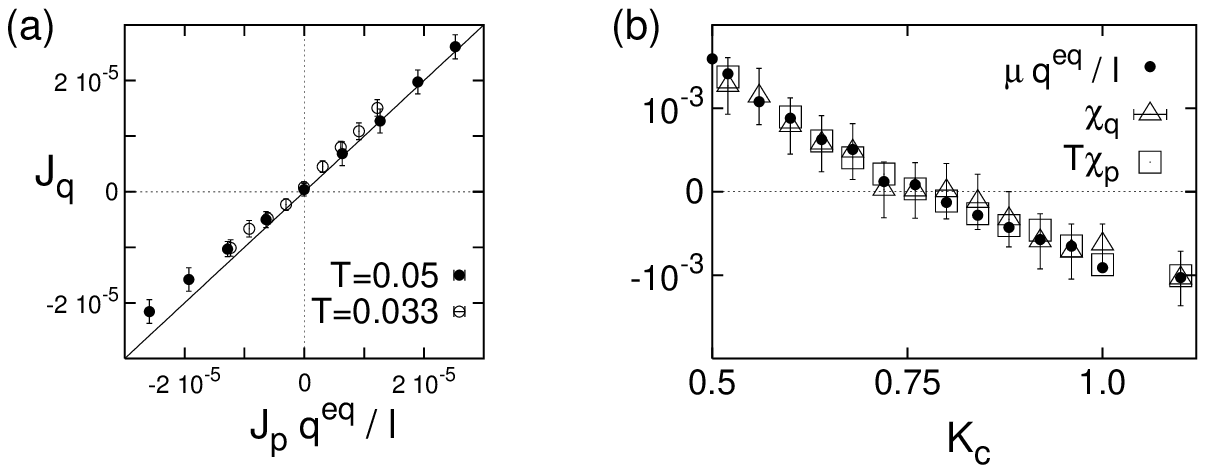}
\end{center}
\caption{
(a) Replot of the data in Fig.\ref{fig:J1,J2}(b).
Both data for $T=0.05$ and $0.033$ are scaled to the same form as 
$J_\mathrm{q}\simeq q^\mathrm{eq} J_\mathrm{p}/l$,
which corresponds to $q_\mathrm{c}\simeq q^\mathrm{eq}$.
$J_\mathrm{q}$ and $J_\mathrm{p}$ are calculated for $f\neq 0$
while $q^\mathrm{eq}$ is evaluated for $f=0$.
(b) The agreement of $\chi_\mathrm{q}$ with $\mu q^\mathrm{eq}/l$
 over various $K_\mathrm{c}$. $T=0.05$.
$\chi_\mathrm{q}\equiv J_\mathrm{q}/f$ and $\mu\equiv J_\mathrm{p}/f$ for small values of $f$
with $\Delta T=0$.
$q^\mathrm{eq}$ is evaluated for $f=0$ and $\Delta T=0$.
At the same time $T\chi_\mathrm{p}$ is plotted,
 where $\chi_\mathrm{p}\equiv J_\mathrm{p}/\Delta T$
 for small values of $\Delta T$ with $f=0$.
The observed equality $\chi_\mathrm{q}=T\chi_\mathrm{p}$
indicates that the reciprocity relation $L_\mathrm{qp}=L_\mathrm{pq}$ holds.
}
\label{fig:Jq-Jp}
\end{figure}

Next, consider the case where a small external force $f$ is applied
 under the iso-thermal condition $\Delta T=0$.
From the measurement of $J_\mathrm{p}$ and $J_\mathrm{q}$,
 we can define a characteristic heat
 $q_\mathrm{c}\equiv J_\mathrm{q}/(J_\mathrm{p}/l)$
 in the linear region of $J_\mathrm{p}$ and $J_\mathrm{q}$.
Because $J_\mathrm{p}/l$ is the net number of directed steps per unit time,
 $q_\mathrm{c}$ is conjectured to be a heat transfered per
 single directed step.
By determining the value of $q_\mathrm{c}$ and $q^\mathrm{eq}$ separately
 from numerical simulation, we have observed
 $q_\mathrm{c}\simeq q^\mathrm{eq}$ for sufficiently small $f$
 (see Fig. \ref{fig:Jq-Jp}(a)).
This implies that the characteristic heat $q_\mathrm{c}$
 associated with the stepwise motion in the linear region
 is the characteristic heat $q^\mathrm{eq}$ at the equilibrium states.
Then, we obtain
\begin{equation}
J_\mathrm{q}=\frac{\mu q^\mathrm{eq}}{l} f
=\frac{q^\mathrm{eq} D_\mathrm{p}}{l} \frac{f}{T},
\label{eqn:chi_q-q}
\end{equation}
by substituting $J_\mathrm{p}=\mu f$ into $J_\mathrm{q}/(J_\mathrm{p}/l)=q^\mathrm{eq}$.
Eq. (\ref{eqn:chi_q-q}) holds for a wide range of the model parameter $K_\mathrm{c}$
as is examined in Fig. \ref{fig:Jq-Jp}(b),
where $\chi_\mathrm{q}\equiv J_\mathrm{q}/f$.
Thus the non-diagonal coefficient $L_\mathrm{qp}$ is obtained
as ${q^\mathrm{eq}D_\mathrm{p}}/{l}$.
In addition, another non-diagonal coefficient $L_\mathrm{pq}$
brings the same form with $L_\mathrm{qp}$
via the agreement $\chi_\mathrm{q}=T\chi_\mathrm{p}$
in Fig. \ref{fig:Jq-Jp}(b),
where $\chi_\mathrm{p}\equiv J_\mathrm{p}/\Delta T$
 for small $\Delta T$ with $f=0$.
We can rewrite the reciprocity coefficients as
 $L_\mathrm{qp}=L_\mathrm{pq}=q^\mathrm{eq}\kappa^\mathrm{eq}l$
 using the rate constant $\kappa^{\mathrm{eq}}$
 for the barrier-overcoming process of the stepwise motion in equilibrium,
 where $D_\mathrm{p}$=$\kappa^\mathrm{eq} l^2$.
This form clearly demonstrates that the characteristic heat brought by the
barrier-overcoming process results in the reciprocity coefficients.

\section{Generality of the form for reciprocity coefficients}

In this section, we show that the expressions (\ref{eqn:reciprocal})
 of the reciprocity coefficients are valid
 in a class of Feynman's ratchet attached to two heat baths.
In this class of systems, we can generally define 
particle flow $J_\mathrm{p}$ and heat flow $J_\mathrm{q}$ between the two baths 
and those temporal quantities $\dot x_\mathrm{p}(t)$ and $\hat J_\mathrm{q}(t)$.
The reciprocity coefficients 
are written with
 the time correlation function in equilibrium states \cite{Kubo} as
\begin{equation}
L_\mathrm{pq} = L_\mathrm{qp} = \lim_{t_0\rightarrow\infty}
\frac{1}{2t_0}
\int_{0}^{t_0}\!\!\!\!\!\mathrm{d}t
\int_{-\infty}^{\infty}\!\!\!\!\!\!\!\mathrm{d}t'
~\langle\dot{x}_\mathrm{p}(t)\hat{J_\mathrm{q}}(t+t')\rangle
\label{eqn:Kubo}
\end{equation}
 where $\langle\cdot\rangle$ denotes ensemble average.
Nonvanishing contribution in the r.h.s. of Eq. (\ref{eqn:Kubo})
 comes from the stepwise motion of the particle.
Supposing that the correlation decays within a finite time and
that each stepwise motion is sufficiently separated in time,
 Eq. (\ref{eqn:Kubo}) is rewritten as a summation of the contribution
 from each stepwise motion at $t=t_{i_\mathrm{s}}$ as
\begin{equation}
\lim_{t_0\rightarrow\infty}\frac{1}{2t_0}\!\!
\sum_{\{i_\mathrm{s} | t_{i_\mathrm{s}}\in (0,t_0)\}}\!\!
\int_{t_{i_\mathrm{s}}-\tau}^{t_{i_\mathrm{s}}+\tau}\!\!\!\!\!\!\mathrm{d}t
~\langle\dot{x}_\mathrm{p}(t)
\int_{-\tau}^{\tau}\!\!\!\!\!\mathrm{d}t'
~\hat{J_\mathrm{q}}(t+t')\rangle,
\label{eqn:Kubo2}
\end{equation}
where $i_\mathrm{s}$ is the index of the stepwise motion
and $\tau$ is a time sufficiently longer than the correlation time.
Because the integration
$\hat{q}_{i_\mathrm{s}} \equiv
\int_{-\tau}^{\tau}\!\mathrm{d}t'\hat{J}_\mathrm{q}(t+t')$
has no explicit dependence on $t$, the integration of $\dot{x}_\mathrm{p}$
with $t$ can be done separately, which
gives $+l$ and $-l$ for rightward and leftward stepwise motion, respectively.
Classifying the stepwise motion indexed by $i_\mathrm{s}$ into the
 rightward one indexed by $i_\mathrm{R}$
and the leftward one indexed by $i_\mathrm{L}$,
Eq. (\ref{eqn:Kubo2}) gives
\begin{equation}
L_\mathrm{pq}\!\! =\!\!
\lim_{t_0\rightarrow\infty}\frac{l}{2t_0}\!\!
\left( 
\sum_{\{i_\mathrm{R} | t_{i_\mathrm{R}}\in (0,t_0)\}} 
\!\!\!\!\!\!\langle \hat{q}_{i_\mathrm{R}}\rangle 
\;\;-\!\!\!\!\!\sum_{\{i_\mathrm{L} | t_{i_\mathrm{L}}\in (0,t_0)\}} 
\!\!\!\!\!\!\langle \hat{q}_{i_\mathrm{L}}\rangle
\right)
=\!\! \frac{l}{2} \left(
\kappa_\mathrm{R}
\langle\hat{q}_{i_\mathrm{R}}\rangle
- \kappa_\mathrm{L}
\langle\hat{q}_{i_\mathrm{L}}\rangle
\right)
= {q^\mathrm{eq} \kappa^\mathrm{eq} ~l}
\label{eqn:Kubo3}
\end{equation}
where 
$\langle \hat{q}_{i_\mathrm{R}}\rangle=q_\mathrm{R}=q^\mathrm{eq}$ and 
$\langle \hat{q}_{i_\mathrm{L}}\rangle=q_\mathrm{L}=-q^\mathrm{eq}$
\cite{Komatsu_Nakagawa}.
$\kappa_\mathrm{R}$ and $\kappa_\mathrm{L}$ are
rate constants for the rightward and the leftward stepwise motion respectively
and in equilibrium 
$\kappa_\mathrm{R}=\kappa_\mathrm{L}=\kappa^\mathrm{eq}$.
The expression of $L_\mathrm{pq}$ in Eq. (\ref{eqn:Kubo3}) is equivalent to
Eq. (\ref{eqn:reciprocal}).
Furthermore, direct numerical calculations of the time integral of
 Eq. (\ref{eqn:Kubo}) shows a good agreement
 with $q^\mathrm{eq}D_\mathrm{p}/l$ 
(data not shown).

\section{Qualification of the heat pump}

\begin{figure}[t]
\begin{center}
\onefigure[scale=1]{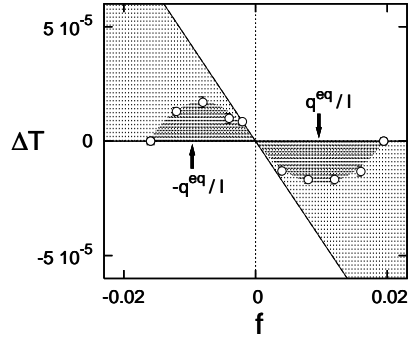}
\end{center}
\caption{
The dark shaded area shows the range where the system 
 functions as a heat pump when $T=0.05$.
Numerically identified boundaries of the second inequality 
of (\ref{eqn:pump}) are depicted by ($\circ$).
The bright shaded area is obtained from Eq. (\ref{eqn:first-order}),
which is the approximation valid in the linear response regime.
}
\label{fig:f-deltaT}
\end{figure}

A system functions as a heat pump
if it can absorb energy from a cooler heat bath
and dissipate it to a hotter one.
The present system can function as a heat pump when the conditions
\begin{equation}
\Delta T J_\mathrm{q} \le 0, \quad  |J_\mathrm{q}|\ge \dot W/2 \ge 0
\label{eqn:pump}
\end{equation}
are valid.
The former inequality means that the heat flow $J_\mathrm{q}$ between the two heat baths
is against temperature gradient $\Delta T$.
The latter means that 
energy is absorbed from one of the heat bath while 
dissipated to the other, i.e., $J_1 J_2 \le 0$
where $J_1=-J_\mathrm{q}-\dot W/2$ and $J_2=J_\mathrm{q}-\dot W/2$.

For the linear response region with small $f$ and $\Delta T$,
the conditions (\ref{eqn:pump}) are reduced to
\begin{equation}
\Delta T (\Delta T+\mu q^\mathrm{eq}f/\lambda l)\le 0,
\label{eqn:first-order}
\end{equation}
which is satisfied in the bright shaded area 
in $f$-$\Delta T$ plane of Fig. \ref{fig:f-deltaT}.
Here we substituted $J_\mathrm{q}$ of
 Eq. (\ref{eqn:reciprocal0}) and (\ref{eqn:reciprocal}) 
into the first inequality of (\ref{eqn:pump}).
$\dot W$ is vanishing in the linear order of 
$f$ and $\Delta T$ because $\dot W=J_p f$,
which means that the second inequality in (\ref{eqn:pump}) always holds.

For larger values of $f$ and $\Delta T$, nonlinear effects
(e.g. nonvanishing $\dot{W}$) are not negligible
 and the region where the system can work as a heat pump
 is reduced.
For instance, for large values of $f$ with $\Delta T=0$,
 we obtain $J_1<0$ and $J_2<0$ as is seen in Fig. \ref{fig:J1,J2}(a)
 although Eq. (\ref{eqn:first-order}) is satisfied.
The region of $f$ and $\Delta T$ satisfying the conditions (\ref{eqn:pump}) 
where the system actually functions as a heat pump are 
numerically identified in the model of Fig. \ref{fig:model}(a)
as the dark shaded area in Fig. \ref{fig:f-deltaT}.
Adjusting the strength of the applied force 
within this confined area of parameters,
we can choose the energy absorption rate from a cooler heat bath.
The rate is maximum 
 at $|f| \simeq |q^\mathrm{eq}|/l$ in the present system 
(refer Fig. \ref{fig:f-deltaT})
\footnote{
In the present model,
Eq. (\ref{eqn:reciprocal0}) and
$\dot W=J_\mathrm{p}f$ give a good estimation of $J_1$ and $J_2$,
which result in maximum energy absorption rate from a cooler heat bath 
at $|f|=|q^\mathrm{eq}|/l(1+|\Delta T|/2T)\simeq |q^\mathrm{eq}|/l=L_{\mathrm{qp}}/L_{\mathrm{pp}}$.
}.
The direction of the heat transfer can also be controlled
 by the direction of the applied force.
The rightward mechanical force
 induces the absorption of heat from $B_2$ and
 the dissipation to $B_1$ (i.e. $J_1\le 0$ and $J_2\ge 0$),
 and vice versa for the leftward force
 ($J_1\ge 0$ and $J_2\le 0$).

With this heat pump, we can cool an object
 to the temperature limited by the conditions (\ref{eqn:pump})
optimizing the applied force.
If we need to achieve lower temperature,
we should construct a proper structure combining multiple pumps.

\section{Discussion}

In this paper, we have shown that heat transfer can be controlled
 by a mechanical force.
We clarified the conditions for the mesoscopic system to function as a heat pump 
based on the study of the linear response regime
 and numerical simulations in fully nonequilibrium.
These results are consistent with the results obtained in a discretized
 solvable model for Feynman's ratchet \cite{Jarzynski_Mazonka}.
The Langevin dynamics adopted in this study
 suggests an intuitive picture based on the barrier overcoming processes
 and associated heat transfer $q^\mathrm{eq}$.
The form of the reciprocity coefficients in Eq. (\ref{eqn:Kubo3})
 generally applies to the class of systems 
with an asymmetric periodic structure and two heat baths. 
When the direct measurement of ${q^\mathrm{eq}}$ is difficult,
 we can estimate ${q^\mathrm{eq}}$ from the Onsager coefficients
 as ${q^\mathrm{eq}}/{l}=L_{\mathrm{qp}}/L_{\mathrm{pp}}$.
This estimation can be applied to other models of heat pump,
 e.g. \cite{Jarzynski_Mazonka} and \cite{Broeck_Kawai}.

The form of the reciprocity coefficient
 $L_\mathrm{qp}=q^\mathrm{eq}\kappa^\mathrm{eq}l$ implies that 
the present heat pump works effectively in small energy scales
where barrier-overcoming processes are governed by thermal fluctuations. 
This is typically realized in molecular scales.
The pump is not efficient in very low temperatures where
the barrier-overcoming event rarely occurs.
It would also be less effective for very high temperatures.
This is because the value of $q^\mathrm{eq}$, 
which depends on the left-right asymmetry of the system,
would be smaller in higher temperatures.

\noindent
{\bf Acknowledgment: }
We are grateful to 
H. Tasaki for a critical reading of this manuscript.
This work is supported by MEXT KAKENHI (No. 16740217).

\noindent
{\bf Note: }
After the completion of the present work, 
\cite{Ai_Wang_Liu} and \cite{Segal_Nitzan} were published 
and \cite{Broeck_Kawai} appeared in the archive,
where different models of heat pump are discussed.


\begin{thebibliography}{999}


\bibitem{Peyrard_Casatti} 
{M. Terraneo, M. Peyrard and G. Casati}:
{Phys. Rev. Lett.}
\textbf{88}
{(2002)}
{094302}.

\bibitem{Humphrey_Linke}
{T. E. Humphrey and H. Linke}:
{Phys. Rev. Lett.}
\textbf{94}
{(2005)}
{096601}.

\bibitem{Feldmann_Kosloff}
{T. Feldmann and R. Kosloff}:
{Phys. Rev. E}
\textbf{61}
{(2000)}
{4774-4790}.

\bibitem{Jarzynski_Mazonka}
{C. Jarzynski and O. Mazonka}:
{Phys. Rev. E}
\textbf{59}
{(1999)}
{6448-6459}.

\bibitem{Feynman}
{R. P. Feynman}:
{Lectures in Physics, Vol.I}
({Addison-Wisley Publishing Co.},{1963}

\bibitem{Nakagawa_Komatsu}
{N. Nakagawa and T. S. Komatsu}:
{J. Phys. Soc. Jpn.}
\textbf{74}
{(2005)}
{1653}.
{Physica A}
\textbf{361}
{(2006)}
{216}.

\bibitem{Sekimoto}
{K. Sekimoto}:
{J. Phys. Soc. Jpn.}
\textbf{66}
{(1997)}
{1234-1237}.

\bibitem{Sekimoto_Takagi_Hondou}
{K. Sekimoto, F. Takagi and T. Hondou}:
{Phys. Rev. E}
\textbf{62}
{(2000)}
{7759-7768}.


\bibitem{Onsager}
{L. Onsager}:
{Phys. Rev.}
\textbf{37}
{(1931)}
{405-426}.
{Phys. Rev.}
\textbf{38}
{(1931)}
{2265-2279}.

\bibitem{Komatsu_Nakagawa}
{T. S. Komatsu and N. Nakagawa}:
{cond-mat/0510838}.


\bibitem{Kubo}
{R. Kubo, M. Toda and N. Hashitsume}:
{Statistical Physics II: Noneqilibrium Statical Mechanics}
{(Springer-Verlag, Berlin, 1991)}.

\bibitem{Broeck_Kawai}
{C. Van den Broeck and R. Kawai}:
{cond-mat/0602153}.

\bibitem{Ai_Wang_Liu}
{B.-Q. Ai, L. Wang and L.-G. Liu}:
{Phys. Lett. A}
\textbf{352}
{(2006)}
{286-290}.

\bibitem{Segal_Nitzan}
{D. Segal and A. Nitzan}:
{Phys. Rev. E}
\textbf{73}
{(2006)}
{026109}.




\end{thebibliography}
\end{document}